\documentclass[onecolumn,useAMS,usenatbib]{mnras}
\usepackage{graphicx}
\usepackage{subfigure}
\usepackage{xcolor}
\usepackage{mathtext,bbm,amsmath,amsfonts,amssymb,indentfirst,syntonly,graphicx}
\usepackage{slashbox}
\usepackage[english]{babel}
\usepackage{calc}
\usepackage{tikz}
\usepackage{bm}
\usepackage{epstopdf}
\usepackage{bm}
\usepackage[utf8]{inputenc}

\def\bc{\begin{center}}
\def\ec{\end{center}}
\def\be{\begin{eqnarray}}
\def\ee{\end{eqnarray}}

\title[Reconstructing the Hubble diagram of GRBs]{Reconstructing the Hubble diagram of gamma-ray bursts using deep learning}
\author[L. Tang, H.-N. Lin, X. Li and L. Liu]
        {Li Tang$^{1,2}$,
        Hai-Nan Lin$^{2}$\thanks{corresponding author: linhn@cqu.edu.cn},
        Xin Li$^{2}$ and
        Liang Liu$^{1,2}$\\
$^{1}$Department of Math and Physics, Mianyang Normal University, Mianyang 621000, China\\
$^{2}$Department of Physics, Chongqing University, Chongqing 401331, China\\}

\begin{document}

\date{Accepted xxxx; Received xxxx; in original form xxxx}

\pagerange{\pageref{firstpage}--\pageref{lastpage}} \pubyear{2021}

\maketitle

\label{firstpage}

\begin{abstract}
  We calibrate the distance and reconstruct the Hubble diagram of gamma-ray bursts (GRBs) using deep learning. We construct an artificial neural network, which combines the recurrent neural network and Bayesian neural network, and train the network using the Pantheon compilation of type-Ia supernovae. The trained network is used to calibrate the distance of 174 GRBs based on the Combo-relation. We verify that there is no evident redshift evolution of Combo-relation, and obtain the slope and intercept parameters, $\gamma=0.856^{+0.083}_{-0.078}$ and $\log A=49.661^{+0.199}_{-0.217}$, with an intrinsic scatter $\sigma_{\rm int}=0.228^{+0.041}_{-0.040}$. Our calibrating method is independent of cosmological model, thus the calibrated GRBs can be directly used to constrain cosmological parameters. It is shown that GRBs alone can tightly constrain the $\Lambda$CDM model, with $\Omega_{\rm M}=0.280^{+0.049}_{-0.057}$. However, the constraint on the $\omega$CDM model is relatively looser, with $\Omega_{\rm M}=0.345^{+0.059}_{-0.060}$ and $\omega<-1.414$. The combination of GRBs and Pantheon can tightly constrain the $\omega$CDM model, with $\Omega_{\rm M}=0.336^{+0.055}_{-0.050}$ and $\omega=-1.141^{+0.156}_{-0.135}$.
\end{abstract}

\begin{keywords}
gamma-ray burst: general  --  cosmology: observations  --  distance scale.
\end{keywords}

\section{Introduction}\label{intro}

According to the standard cosmological model ($\Lambda$CDM model), the Universe is composed of radiation, baryon, dark matter and dark energy. The radiation is negligible, while the dark matter and dark energy dominate at present day. The dark energy provides a negative pressure to equilibrate the action of gravity and results in the accelerating expansion of the Universe \citep{Mortonson et al.:2013,Li et al.:2013,Arun et al.:2017,Brax:2018}. Many observations have confirmed the accelerated expanding universe, such as type-Ia supernovae (SNe Ia) \citep{Riess:1998,Perlmutter:1999}, baryon acoustic oscillations (BAO) \citep{Eisenstein:2005,Alam:2017}, the strong gravitational lensing time delays \citep{Suyu:2013}, cosmic microwave background (CMB) radiation \citep{Planck Collaboration:2014,Planck Collaboration:2020}, and so on. The local probes such as SNe Ia and BAO are detectable only at low redshift $z\lesssim 2$, thus they can be used to investigate the local Universe. On the other hand, CMB probes the Universe at the time of photon decoupling at redshift $z\approx 1089$. There is a big redshift gap between the local probes and CMB. Recently, the tension of the Hubble constant measured from local data and measured from CMB arises great challenges on the standard cosmological model \citep{Riess:2016fkd,Riess:2018uxu,Verde:2019}. Gamma-ray bursts (GRBs) provide a supplementary tool to investigate the Universe in this redshift gap.

GRBs are the most energetic explosions in the Universe, whose isotropic equivalent energy released in a few seconds can be up to 10$^{55}$ erg, see e.g. \citep{Kumar and Zhang:2015,Zhang:2018,Levan:2018} for recent review. The high luminosity makes GRBs to be detectable to high redshift, e.g., the furthest GRB observed hitherto is GRB 090429B at redshift $z=9.4$ \citep{Cucchiara et al.:2011}. Therefore, GRBs hopefully become the complementary probes to fill the redshift gap between SNe Ia and CMB. Comparing with SNe Ia, GRBs are much less affected by the extinction of interstellar medium. Unfortunately, they are hard to calibrate due to their complex triggering mechanisms. Nevertheless, many empirical luminosity correlations between various observables in both the prompt and afterglow emission phases are found \citep{Fenimore:2000,Norris:2000,Amati:2002,Yonetoku:2004,Ghirlanda et al.:2004,Schaefer:2007,Izzo et al.:2015}. Based on these luminosity correlations, several methods have been proposed to standardize GRBs so that they can be employed to constrain cosmological parameters \citep{Dai et al.:2004,Firmani:2005,Liang:2008,Wei:2009,Izzo et al.:2009,Wei:2010,Liu:2015,Lin:2015fma}.

It is well known that when calibrating GRBs one usually encounters the ``circularity" problem. This is because all the luminosity correlations depend on the luminosity distance, which depends on the cosmological model. In other words, in absence of a primary distance calibrator for GRB relations, one must rely on secondary distance indicators such as SNe Ia, or taking a prior distance based on a specific cosmological model. In the later case, the distance is cosmologically model-dependent. Thus, the distance calibrated in one cosmological model couldn't be directly used to constrain the other models. To avoid the circularity problem, several model-independent methods are proposed \citep{Liang:2008,Wei:2009,Wei:2010,Liu:2015}. The main procedures of calibration are: (i) dividing GRBs into low-$z$ and high-$z$ subsamples; (ii) obtaining the luminosity distance of low-$z$ GRBs according to the distance-redshift relation derived from e.g. SNe Ia; (iii) calibrating the luminosity correlations, e.g. the Amati relation, of low-$z$ GRBs; (iv) assuming that high-$z$ GRBs follow the same correlation, one can inversely calculate the luminosity distance of high-$z$ GRBs. These calibrating methods are based on the assumption that the luminosity correlations are not evolving with redshift. However, several works indicate that some luminosity correlations may not be universal over all redshifts \citep{Li:2007,Wang et al.:2011,Lin:2015,Lin:2015fma,Tang et al.:2021}.

Recently, \citet{Izzo et al.:2015} proposed a new GRB correlation, namely the Combo-relation, and calibrated the relation with an alternative method. They first inferred the slop parameter $\gamma$ with a small but sufficient subsample of GRBs approximately at the same redshift. Fixing the slop parameter, they determined the intercept parameter $A$ by comparing the nearest GRB with five SNe Ia located at the very similar redshift. After obtaining the parameters of the Combo-relation, \citet{Izzo et al.:2015} calibrated 60 GRBs whose highest redshift is up to $z=8.2$. \citet{Muccino et al.:2021} applied the similar method to a much larger sample of GRBs and verified that the slope parameter of Combo-relation is redshift-independent. One shortcoming of this method is that, at high redshift it is difficult to find sufficient GRBs approximately at the same redshift. The slope parameter $\gamma$ is obtain from a small subsample, not the full sample of GRBs. The intercept parameter $A$ is determined only from a handful of SNe Ia located at the same redshift of the nearest GRBs. Thus, although this method can minimize the use of SNe Ia, it is strongly dependents on the precision of the few SNe used.

Based on the Combo-relation, we will show in this paper that GRBs can be calibrated using  deep learning method. Deep learning is a subset of machine learning in artificial intelligence that imitates the workings of human brain in processing information. It can deal with some complex work which is hard to do using traditional methods, such as image classification, signal identification, machine translation. With the tremendous development of computer technology, deep learning has been developed rapidly recent years and it has been widely used in various fields, such as data mining, natural language processing, security, traffic domain, and so on. Recently, deep learning has been successfully applied in cosmology, such as predicting galaxy morphology \citep{Dieleman:2015}, constraining dark energy \citep{Escamilla-Rivera et al.:2020}, calibrating GRBs \citep{Luongo:2020hyk}, etc. In a recently paper \citep{Tang et al.:2021}, we have used deep learning to investigate the redshift dependence of six luminosity correlations of GRBs. We have constructed a network combining the Recurrent Neural Network (RNN) and the Bayesian Neural Network (BNN), and trained the network with supervovae. Using the trained network, we studied six luminosity correlations and found that only one out of the six correlations (namely the $E_p-E_\gamma$ relation) has no evidence for redshift dependence. In this paper, the same network is applied to the Combo-relation. We divide 174 GRBs into low-$z$ and high-$z$ subsamples with the redshift boundary $z=2$. The network is used to calibrate the Combo-relation for both low-$z$ and high-$z$ subsamples, as well as for the full sample, to see if there is any redshift dependence. Our method can not only test the possible redshift dependence of the Combo-relation model independently, but also can obtain the slope and intercept parameters simultaneously for the full GRB sample. Therefore, we can reconstruct the Hubble diagram of the full GRB sample and use it to constrain cosmological models.

The rest of this paper is organized as follows: In Section \ref{sec:calibration}, we introduce the network and use it to test the possible redshift dependence of Combo-relation. In Section \ref{sec:cosmology}, we calibrate the distance of GRBs and show its application in constraining cosmological models. Finally, discussions and conclusions are presented in Section \ref{sec:conclusion}.

\section{testing the redshift dependence of Combo-relation}\label{sec:calibration}

Combining the well-known Amati relation \citep{Amati:2002}, which is a correlation between the peak energy of the $\nu F_\nu$ spectrum and isotropic equivalent energy, and the results of \citet{Bernardini et al.:2012} and \citet{Margutti et al.:2013}, which relates the X-ray and $\gamma$-ray isotropic energies to the peak energy, \citet{Izzo et al.:2015} achieved the Combo-relation,
\begin{equation}\label{eq:Combo}
\log\left(\frac{L_0}{\rm erg/s}\right)=
\log\left(\frac{A}{\rm erg/s}\right)+
\gamma\log\left(\frac{E_{p,i}}{\rm keV}\right)-
\log\left(\frac{\tau/\rm s}{|1+\alpha|}\right),
\end{equation}
where $\log$ is the logarithm of base 10, $L_0$ is the isotropic equivalent luminosity at plateau, $\gamma$ and $A$ are respectively the slope and intercept parameters, $E_{p,i}$ is the rest-frame peak energy of the $\nu F_{\nu}$ spectrum, $\alpha$ and $\tau$ are the late power-law decay index and the characteristic timescale of the end of plateau, respectively. Given the luminosity distance of a GRB, the plateau luminosity $L_0$ can be derived from the rest-frame $0.3-10$ keV energy flux $F_0$ by
\begin{equation}\label{eq:L_dL}
L_0=4\pi d^2_L F_0,
\end{equation}
where the luminosity distance $d_L$ is related to the dimensionless distance modulus $\mu$ by
\begin{equation}\label{eq:mu}
\mu=5\log\frac{d_L}{\rm Mpc}+25.
\end{equation}
The uncertainty of $L_0$ propagates from the uncertainties of $F_0$ and $d_L$,
\begin{equation}\label{eq:L0_error}
\frac{\sigma_{L_0}}{L_0}=\sqrt{\left(\frac{\sigma_{F_0}}{F_0}\right)^2+4\left(\frac{\sigma_{d_L}}{d_L}\right)^2}.
\end{equation}
Defining
\begin{equation}\label{eq:xy}
  y\equiv\log\left(\frac{L_0}{\rm erg/s}\right), a\equiv\log\left(\frac{A}{\rm erg/s}\right), x_1\equiv\log\left(\frac{E_{p,i}}{\rm keV}\right), x_2\equiv\log\left(\frac{\tau/\rm s}{|1+\alpha|}\right),
\end{equation}
The Combo-relation can be simply written as
\begin{equation}
  y=a+\gamma x_1-x_2.
\end{equation}
The uncertainties on $y$, $x_1$ and $x_2$ are given by
\begin{equation}
  \sigma_y=\frac{1}{\ln10}\frac{\sigma_{L_0}}{L_0},\sigma_{x_1}=\frac{1}{\ln10}\frac{\sigma_{E_{p,i}}}{E_{p,i}},
  \sigma_{x_2}=\frac{1}{\ln10}\sqrt{\left(\frac{\sigma_{\tau}}{\tau}\right)^2+\left(\frac{\sigma_\alpha}{1+\alpha}\right)^2}.
\end{equation}

The plateau luminosity $L_0$, therefore the Combo-relation, depends on the luminosity distance, which further depends on the cosmological model. To avoid the model-dependence, we first reconstruct the distance redshift relation from the Pantheon dataset \citep{Scolnic:2018} using deep learning. Deep learning can effectively learn from the complex data to tackle problems by considering Artificial Neural Networks (ANN) as an underlying model, such as Convolutional Neural Networks (CNN), Recurrent Neural Networks (RNN), Bayesian Neural Networks (BNN) and so on. Thereinto, RNN is capable of predicting the future from the complex sequential information, so that we can train RNN with the Pantheon data to predict the trend of distance-redshift relation even beyond the redshift range of data points. However, RNN can not provide the uncertainty of prediction. Fortunately, BNN can fix up this shortcoming. In our work, we introduce the dropout technique into RNN to realize BNN because a traditional BNN is too complex to design. Many works have shown that a network with a dropout, an approximation of the Gaussian process, is equivalent to the Bayesian model \citep{Gal:2016a,Gal:2016b,Gal:2016c,Louizos:2016}. Additionally, dropout is employed to prevent the network from over-fitting, which results from a large number of hyperparameters in the network itself. Therefore, we focus on constructing the RNN with a dropout to reconstruct data, and BNN is achieved when we execute the well-trained network many times.

The details of our network are described in \citet{Tang et al.:2021}. Here we just shortly introduce the structure of the network. There are three layers in our network. The first layer is the input layer to receive the feature (the redshift $z$ here); the second layer called hidden layer is to transform the information from the input layer; and the last layer is the output layer to export the target (distance modulus $\mu$ here). Given any redshift, the network returns the corresponding distance modulus at that redshift. Feeding in the Pantheon data, RNN is trained by minimizing the loss function, which describes the difference between the predictions and the observations. We choose the loss function to be the mean-squared-error (MSE) function in our network, and the Adam optimizer is employed to find its minimum. We set the time step $t=4$ to reduce the training time and adopt the Long Short-Term Memory (LSTM) cell as the basic cell of RNN to avoid the information loss. In order to improve the performance of network, a non-linear function called activation function $A_f$ is introduced. Several activation functions are available, such as the Relu function, Elu function, Selu function, tanh function, and so on. \citet{Tang et al.:2021} have made a detailed comparison between these four activation functions, and found the tanh function performs best in constructing the distance-redshift relation of Pantheon. Therefore, we adopt the tanh as the activation function here, which reads
\begin{equation}
A_f=\tanh(x)=\frac{e^x-e^{-x}}{e^x+e^{-x}}.
\end{equation}
The reliability of the activation function has been checked using Monte Carlo simulation \citep{Tang et al.:2021}.

We train our network using the Pantheon dataset \citep{Scolnic:2018}, which is composed of 1048 SNe Ia in the redshift range from 0.01 to 2.3. The light curve parameters of SNe Ia are transformed into the distance according to a modified version of the Tripp formula \citep{Tripp:1998}
\begin{equation}
\mu=m_B+\alpha x_1-\beta c-M_B+\Delta_M+\Delta_B,
\end{equation}
where $m_B$ is the B-band apparent magnitude, $M_B$ is the absolute magnitude, $x_1$ and $c$ are the stretch and color parameters, $\Delta_M$ is a distance correction based on the host galaxy mass of the SNe, $\Delta_B$ is a distance correction based on predicted biases from simulations, $\alpha$ and $\beta$ are the coefficients of stretch and color corrections, respectively. For the Pantheon sample, \citet{Scolnic:2018} used the BEAMS with Bias Corrections (BBC) method \citep{Kessler and Scolnic:2017} to determine the nuisance parameters $\alpha$ and $\beta$. The Pantheon data set is finally reported as the corrected apparent magnitude $m_{\rm corr}=\mu+M_B$ after correcting the effects of $x$ and $c$. In order to convert the (corrected) apparent magnitude to distance modulus, the absolute magnitude of SNe Ia should be known. Here we fix the absolute magnitude to $M_B = -19.36$ mag \citep{Scolnic et al.:2014}.

\begin{figure}
\centering
\includegraphics[width=0.6\textwidth]{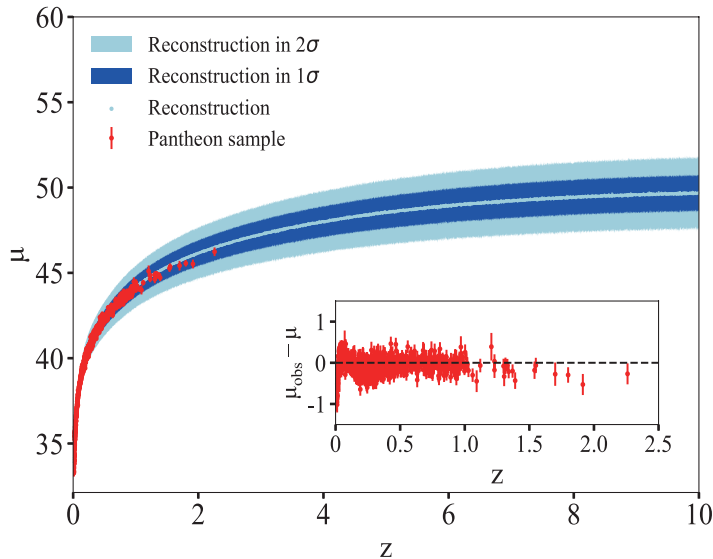}
\caption{\small{The reconstruction of distance-redshift relation up to $z=10$ from Pantheon data set using deep learning. The red dots with $1\sigma$ error bars represent the Pantheon data points. The light-blue dots represent the central values of the reconstruction. The dark blue and light blue regions represent the 1$\sigma$ and 2$\sigma$ uncertainties, respectively. The inset is the residual of data points with respect to the central values of the reconstruction.}}\label{fig_sim}
\end{figure}

The steps of reconstruction and the corresponding hyperparameters of our network are briefly presented as follow (see \citet{Tang et al.:2021} for more details):
(a) Data preprocessing. We first sort the Pantheon data points by redshift from low to high, and normalize the distance moduli. Then we re-arrange $\mu-z$ into four sequences.
(b) Constructing the network. We build RNN with an input layer, a hidden layer and an output layer. The first two layers are built with the LSTM cells of 100 neurons in each. The redshifts and the corresponding distance moduli in each sequence are the input and output vectors, respectively. For the dropout rate, we set it to be 0 in the input layer and 0.2 in the second and the output layers.
(c) Training the network. We train the network 1000 times and save the well-trained network. The distribution of distance-redshift relation is achieved when we execute the network 1000 times. The result is shown in Figure \ref{fig_sim}, where we have reconstructed the distance-redshift relation up to $z=10$. The red dots with $1\sigma$ error bars represent the Pantheon data points. The light-blue dots represent the central values of the reconstruction. The dark blue and light blue regions represent the 1$\sigma$ and 2$\sigma$ uncertainties of the reconstruction, respectively. The inset of Figure \ref{fig_sim} is the residual of data points with respect to the central values of reconstruction. At low redshift, the reconstructed curve is well consistent with the data points. Due to the sparsity of data points at high redshift, the reconstructed central values seem to slightly deviate from the data points beyond $z\gtrsim 1.5$. But in consideration of the reconstructing uncertainty, they are still consistent within 1$\sigma$ confidence level.

After obtaining the distance-redshift relation, we can calculate the distance at any given redshift. We divide the 174 GRBs from \citet{Muccino et al.:2021} into two subsamples, the low-$z$ ($z<2$) sample containing 93 GRBs and the high-$z$ ($z>2$) sample containing 80 GRBs. We exclude GRB 171205A because $\alpha=-1$ is unphysical in equation \ref{eq:Combo}. Then we investigate the redshift-dependence of Combo-relation with this two subsamples as well as the full GRBs sample. The Combo-relation is fitted by maximizing the likelihood \citep{DAgostini:2005}
\begin{equation}
\mathcal{L}\left(a, \gamma, \sigma_{\rm int}\right)\propto\prod_i\frac{1}{\sqrt{\sigma^2_{\rm int}+\sigma^2_y+\gamma^2\sigma^2_{x_1}+\sigma^2_{x_2}}}\times \exp\left[-\frac{(y-a-\gamma x_1+x_2)^2}{2(\sigma^2_{\rm int}+\sigma^2_y+\gamma^2\sigma^2_{x_1}+\sigma^2_{x_2})}\right],
\end{equation}
where we have introduced the intrinsic scatter $\sigma_{\rm int}$, which is regarded as a free parameter, to account for any other unknown errors except for the measurement error.

\begin{figure}\label{fig_xy}
\centering
\includegraphics[width=0.6\textwidth]{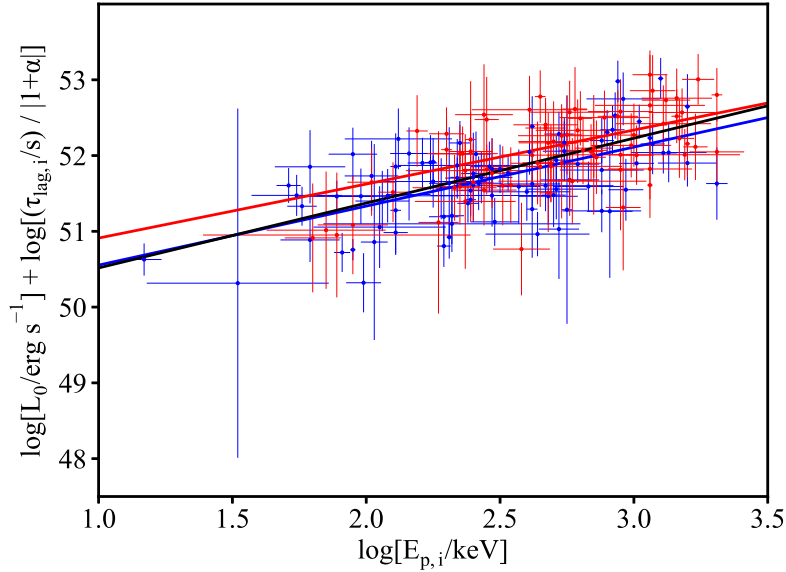}
\caption{\small{The Combo relation for low-$z$ (blue dots) and high-$z$ (red dots) GRBs plotted in the 2-dimensional plane. Error bars denote the 1$\sigma$ uncertainties. The blue, red and black lines are the best-fitting results for low-$z$, high-$z$ and all-$z$ GRBs, respectively.}}
\end{figure}

\begin{figure}
\centering
\includegraphics[width=0.6\textwidth]{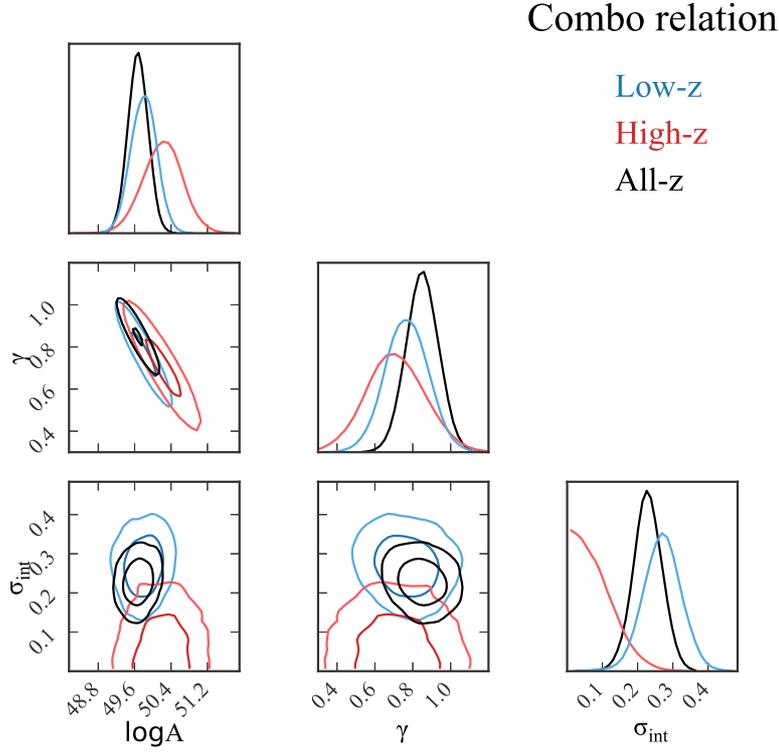}
\caption{\small{The 1-dimensional PDFs and 2-dimensional confidence contours for the parameters of Combo-relation.}}\label{fig_contour}
\end{figure}

Assuming a flat prior on all the free parameters and restrict $\sigma_{\rm int}>0$, we obtain the posterior probability density function (PDF) of the parameter space through Markov Chain Monte Carlo analysis \citep{ForemanMackey:2012ig}. The best-fitting parameters are listed in Table \ref{tab:parameters}, and the Combo-relation in logarithmic coordinates is plotted in Figure \ref{fig_xy}. Low-$z$ and high-$z$ GRBs are denoted with blue and red dots with 1$\sigma$ error bars, respectively. The best-fitting lines for low-$z$ GRBs, high-$z$ GRBs and all-$z$ GRBs are plotted in blue, red and black, respectively. In Figure \ref{fig_contour}, we plot the posterior PDFs, and the 1$\sigma$ and 2$\sigma$ confidence contours of the parameter space.

\begin{table}
\centering
\caption{\small{The best-fitting parameters of the Combo-relation. $N$ is the number of GRBs in each sample.}}\label{tab:parameters}
\arrayrulewidth=1.0pt
\renewcommand{\arraystretch}{1.3}
{\begin{tabular}{c|ccc|c} 
\hline\hline 
Sample & $N$  & $a\equiv\log(A/{\rm erg/s})$     & $\gamma$   &$\sigma_{\rm int}$\\
\hline
low-$z$  &93 &49.775$^{+0.264}_{-0.282}$	&0.779$^{+0.115}_{-0.107}$	 &0.268$^{+0.049}_{-0.050}$\\
high-$z$ &80 &50.198$^{+0.409}_{-0.427}$	&0.713$^{+0.153}_{-0.148}$	 &$<$0.099		\\
All-$z$ &173 &49.661$^{+0.199}_{-0.217}$	&0.856$^{+0.083}_{-0.078}$	 &0.228$^{+0.041}_{-0.040}$\\
\hline
\end{tabular}}
\end{table}

As the results show, the Combo-relation has a small intrinsic scatter for both low-$z$ and high-$z$ GRBs, consistent with the results of \citet{Izzo et al.:2015} and \citet{Muccino et al.:2021}. Especially, for high-$z$ subsample we only obtain the upper limit of intrinsic scatter. Low-$z$ GRBs have a smaller intercept and a sharper slope than high-$z$ GRBs, but the intercept and slope of both subsamples are consistent with each other within 1$\sigma$ uncertainty. This implies that there is no evidence for the redshift evolution of Combo-relation. Therefore, we calibrate the Combo-relation using the full GRB sample, and obtain the best-fitting parameters $a=49.661^{+0.199}_{-0.217}$, $\gamma=0.856^{+0.083}_{-0.078}$ and $\sigma_{\rm int}=0.228^{+0.041}_{-0.040}$. The slope parameter $\gamma$ of the full sample is somewhat larger than that of both the low-$z$ and high-$z$ subsamples, while the intercept parameter $a$ is smaller than that of both subsamples. This is because $\gamma$ and $a$ are highly inversely correlated, as is seen in the contour plot in Figure \ref{fig_contour}. A larger slope will lead to a smaller intercept. Nevertheless, the slope and intercept parameters of low-$z$, high-$z$ and full-$z$ samples are consistent with each other within $1\sigma$ uncertainty.

\section{Distance calibration and constraints on the cosmological models}\label{sec:cosmology}

\begin{figure}
\centering
\includegraphics[width=0.6\textwidth]{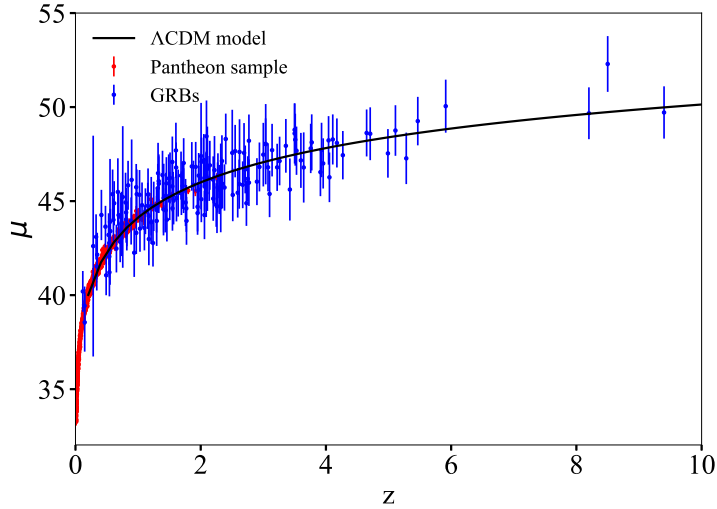}
\caption{\small{The Hubble diagram of GRBs. The red dots with 1$\sigma$ error bars are the Pantheon data points. The blue dots with 1$\sigma$ error bars are the 173 GRBs calibrated with Combo-relation. The black line is the best-fitting result to $\Lambda$CDM model using GRBs only.}} \label{fig_Hubble}
\end{figure}

By fixing the slope and intercept parameters to the best-fitting values, we can calibrate the distance of GRBs using the Combo-relation.
The luminosity distance $d_L$ of GRBs is resolved from equations (\ref{eq:Combo}) and (\ref{eq:L_dL}). Then take the logarithm we obtain the distance modulus of GRBs \citep{Izzo et al.:2015,Muccino et al.:2021},
\begin{equation}\label{eq:mu_GRB}
\mu_{\rm GRB}=-97.45+\frac{5}{2}\left[
\log\left(\frac{A}{\rm erg/s}\right)+
\gamma\log\left(\frac{E_{p,i}}{\rm keV}\right)-
\log\left(\frac{\tau/\rm s}{|1+\alpha|}\right)-
\log\left(\frac{F_0}{\rm erg/cm^2/s}\right)-\log4\pi
\right].
\end{equation}
With the symbols in equation (\ref{eq:xy}), and defining $x_3\equiv \log\left[F_0/({\rm erg/cm^2/s})\right]$, equation (\ref{eq:mu_GRB}) can be rewritten as
\begin{equation}
  \mu_{\rm GRB}=-97.45+2.5(a+\gamma x_1-x_2-x_3-\log 4\pi).
\end{equation}
The uncertainty of $\mu_{\rm GRB}$ is propagated from the uncertainties of the observables $(F_0,E_{p,i},\tau,\alpha)$, as well as that of the Combo-relation parameters $(a,\gamma,\sigma_{\rm int})$,
\begin{equation}
\sigma_{\mu_{\rm GRB}}=2.5\sqrt{\sigma_a^2+x_1^2\sigma_{\gamma}^2+\gamma^2\sigma_{x_1}^2+\sigma_{x_2}^2+\sigma_{x_3}^2+\sigma_{\rm int}^2}.
\end{equation}
The distance of GRBs calibrated using the Combo-relation of the full sample, together with the distance of Pantheon, are plotted in Figure \ref{fig_Hubble}, where the red and blue dots with 1$\sigma$ error bars stand for the Pantheon and GRBs, respectively.

Next, we explore the application of GRBs in constraining cosmological models. In the spatially-flat FRW spacetime, the luminosity distance is given by
\begin{equation}\label{eq:dL}
d_L(z)=(1+z)\frac{c}{H_0}\int^z_0\frac{dz}{E(z)},
\end{equation}
where $c$ is the speed of light, $H_0$ is the Hubble constant at present time, $E(z)\equiv H(z)/H_0$ is the normalized Hubble parameter. In the $\Lambda$CDM model, where the dark energy is a constant with equation-of-state $\omega\equiv p/\rho=-1$, $E(z)$ is given by
\begin{equation}\label{eq:E_LCDM}
E^2(z)=\Omega_{\rm M}(1+z)^3+(1-\Omega_{\rm M}),
\end{equation}
where $\Omega_{\rm M}$ is the matter density at present time. In the $\omega$CDM model where the equation-of-state of dark energy $\omega$ is a free parameter, $E(z)$ is given by
\begin{equation}\label{eq:E_wCDM}
E^2(z)=\Omega_{\rm M}(1+z)^3+(1-\Omega_M)(1+z)^{3(1+\omega)}.
\end{equation}

We use the calibrated GRBs to constraint the parameters of $\Lambda$CDM and $\omega$CDM models. The best-fitting parameters are the ones which can minimize the $\chi^2$,
\begin{equation}
\chi^2_{\rm GRB}=\sum^{N}_{i=1}\left(\frac{\mu_{{\rm GRB},i}-\mu_{\rm th}(\bm\theta;z_i)}{\sigma_{\mu_{{\rm GRB},i}}}\right)^2,
\end{equation}
where $\mu_{\rm GRB}$ is the distance modulus of GRBs, $\mu_{\rm th}$ is the theoretical distance modulus, $\sigma_{\mu_{\rm GRB}}$ is the uncertainty of $\mu_{\rm GRB}$, and $\bm\theta$ is the set of free parameters. Since the calibration of Pantheon has taken a prior $H_0=70 {\rm km~s^{-1}~Mpc^{-1}}$ \citep{Scolnic:2018}, we fix $H_0$ to this value in the fitting. Therefore, there is only one free parameter $\bm\theta=\Omega_M$ in the $\Lambda$CDM model, and two free parameters $\bm\theta=(\Omega_M,\omega)$ in the $\omega$CDM model. We determine the cosmological parameters by maximizing the likelihood function $\mathcal{L}\propto \exp(-\chi^2/2)$, which is equivalent to minimize $\chi^2$. In comparison, we also use the Pantheon dataset to fit cosmological model, and the corresponding $\chi^2$ is given by
\begin{equation}
\chi^2_{\rm SNe}=\Delta\bm\mu^{T}\bm C^{-1}\Delta\bm\mu,
\end{equation}
where the vector of distance residual of SNe Ia is $\Delta\bm\mu=\bm\mu_{\rm SNe}-\bm\mu_{\rm th}(\bm\theta,z)$, and the total covariance matrix $\bm{C}$ is given by
\begin{equation}
\bm{C}=\bm{D}_{\rm stat}+\bm{C}_{\rm sys},
\end{equation}
where $\bm{D}_{\rm stat}$ and $\bm{C}_{\rm sys}$ are respectively the diagonal statistical covariance matrix and the systematic covariance matrix presented in \citet{Scolnic:2018}. Additionally, we also combine the SNe Ia and GRBs to constrain the cosmological parameters, and the total $\chi^2$ is given by $\chi^2_{\rm total}=\chi^2_{\rm GRB}+\chi^2_{\rm SNe}$.

\begin{figure}
\centering
  \includegraphics[width=0.4\textwidth]{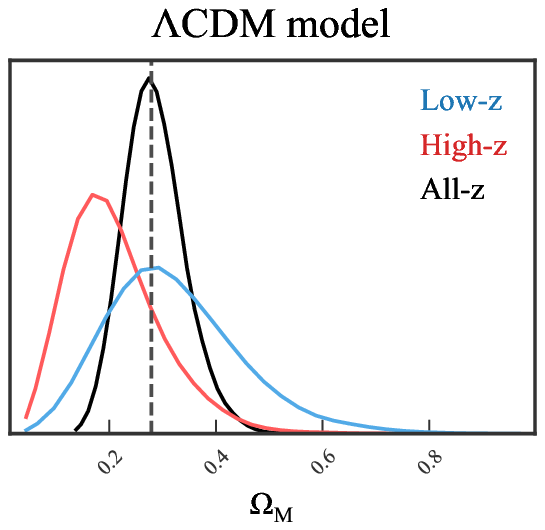}
  \includegraphics[width=0.5\textwidth]{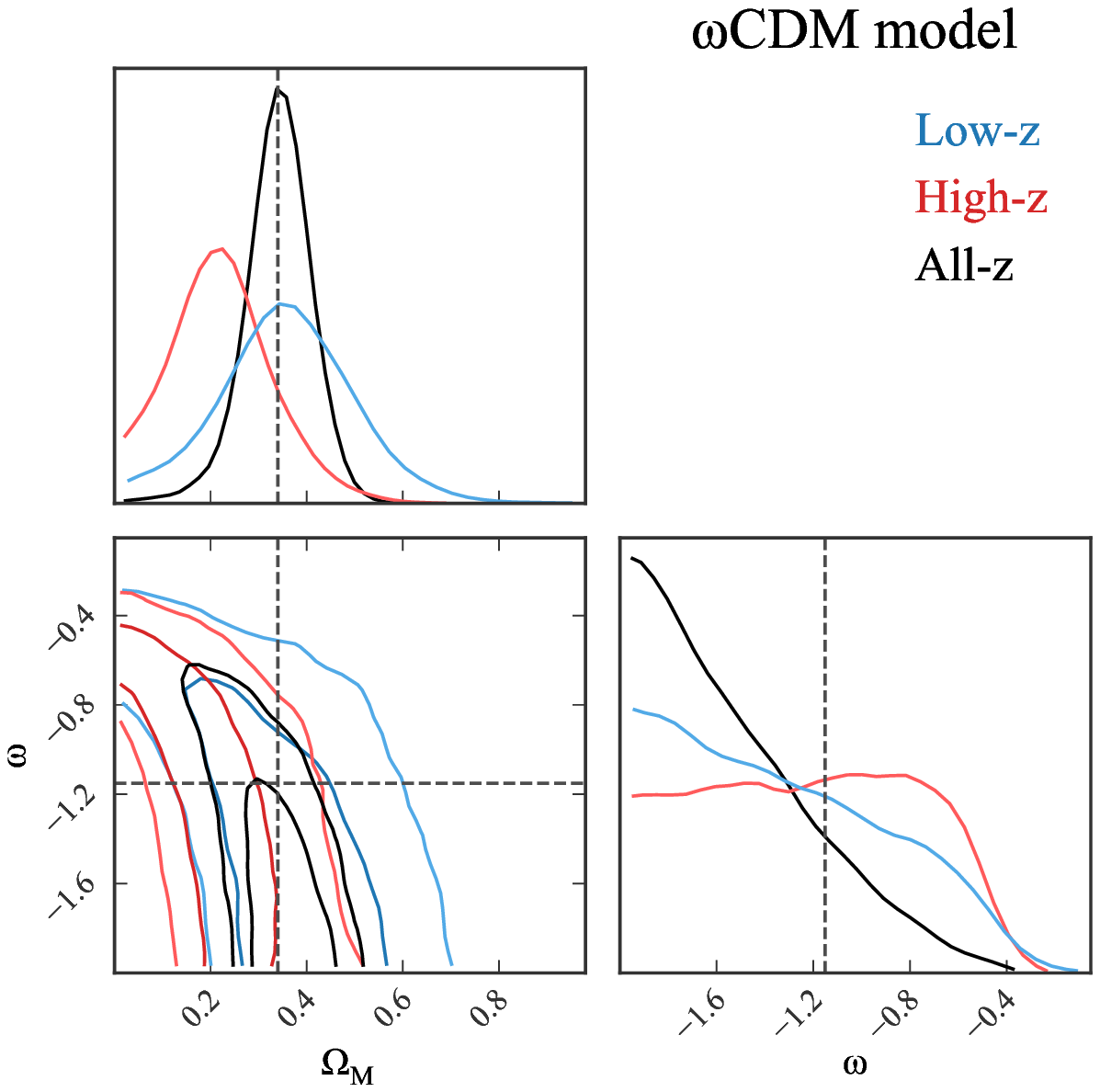}
  \caption{\small{The 1-dimensional PDFs and 2-dimensional confidence contours for the parameters of $\Lambda$CDM model (left) and $\omega$CDM model (right) constrained with different GRB subsamples. The dashed lines are the best-fitting results with Pantheon SNe.}}\label{fig_contour_G}
\end{figure}

\begin{figure}
\centering
  \includegraphics[width=0.4\textwidth]{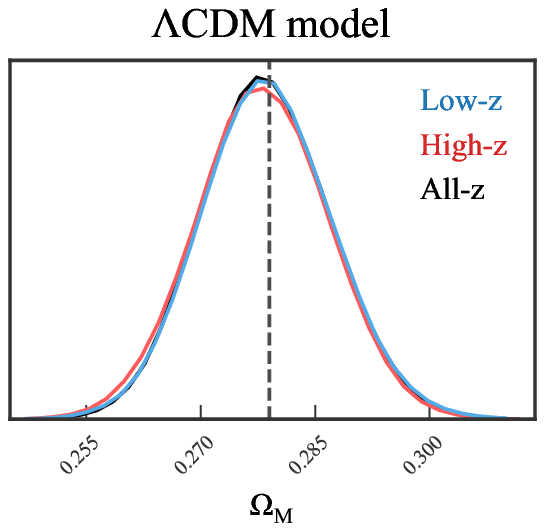}
  \includegraphics[width=0.5\textwidth]{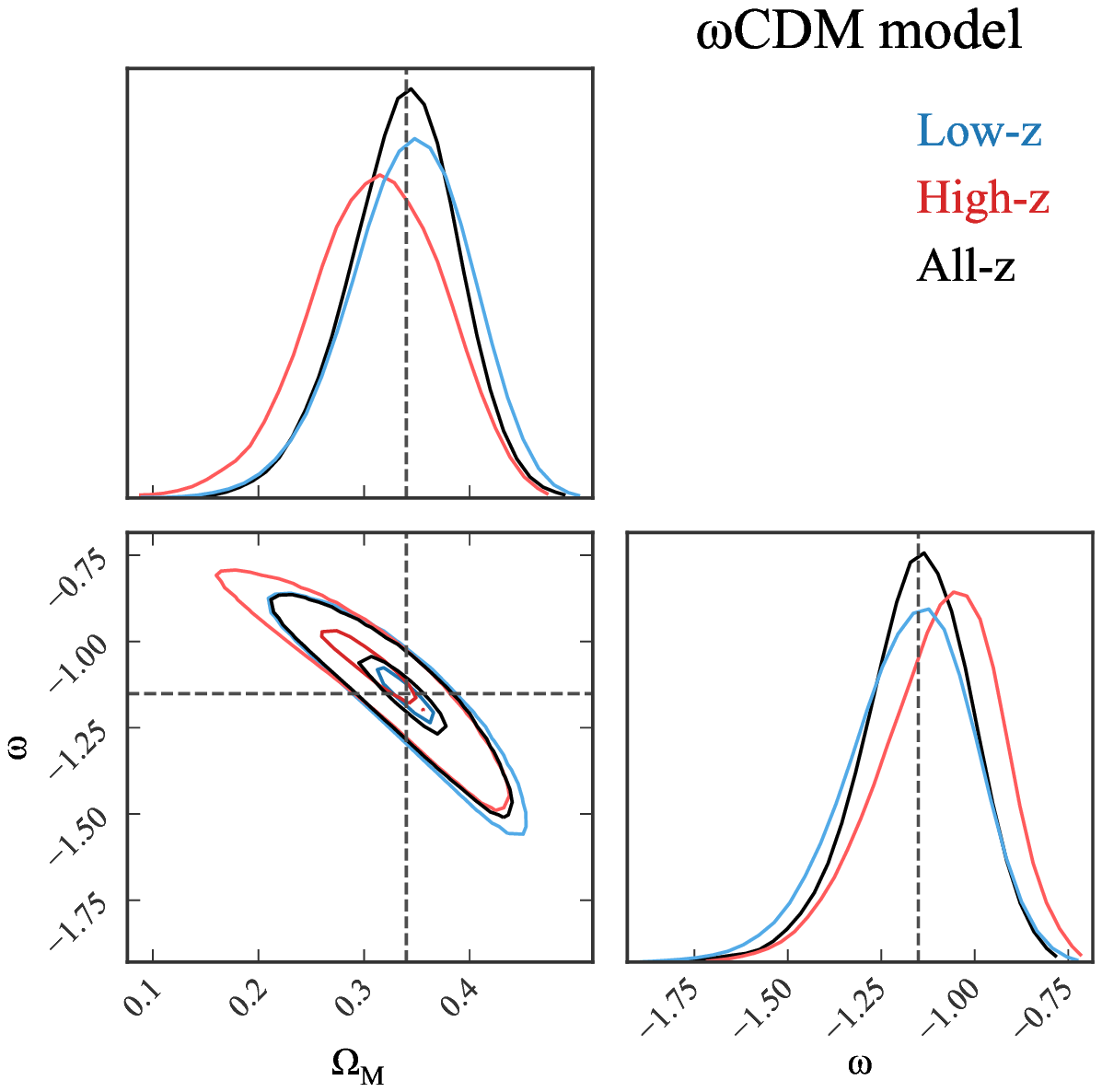}
  \caption{\small{The 1-dimensional PDFs and 2-dimensional confidence contours for the parameters of $\Lambda$CDM model (left) and $\omega$CDM model (right) constrained with the combination of SNe Ia and different GRB subsamples. The dashed lines are the best-fitting results with Pantheon SNe.}}\label{fig_contour_SG}
\end{figure}

For comparison, we also calibrate low-$z$ and high-$z$ subsamples separately. Then these two subsamples are used (or combined with SNe) to constrain the cosmological models. The best-fitting parameters of $\Lambda$CDM model and $\omega$CDM model using GRBs and GRB+SNe are listed in Table \ref{tab:cosmological parameters_G} and \ref{tab:cosmological parameters_SG}, respectively. The corresponding 1-dimensional PDFs and 2-dimensional confidence contours are plotted in Figure \ref{fig_contour_G} and Figure \ref{fig_contour_SG}, respectively. As is seen, for the $\Lambda$CDM model, the constraints on $\Omega_{\rm M}$ with different GRB subsamples are marginally consistent with the result of Pantheon sample ($\Omega_{\rm M}=0.279\pm0.008$) within $1\sigma$ uncertainty. The best-fitting $\Omega_{\rm M}$ from high-$z$ subsample is somewhat smaller than the Pantheon result due to the bias of the reconstruction at high redshift. For the $\omega$CDM model, the Pantheon sample constrains the parameters with $\Omega_{\rm M}=0.340^{+0.063}_{-0.054}$ and $\omega=-1.151^{+0.174}_{-0.153}$. While with GRB subsamples alone, the constraints are loose. Especially with the low-$z$ subsample and all-$z$ sample, the results of constrains on $\omega$ couldn't exclude the possible evolution of EoS of dark energy. Compared with SNe, the GRB sample is so small that it only provides an upper limit of $\omega$. The combination of GRBs with SNe can tightly constrain $\omega$. The constrains on $\Lambda$CDM with a combination of Pantheon sample and different GRB subsamples are consistent with each other, as well as with the results of Pantheon sample alone. This is because the number of GRBs is much smaller than that of SNe, hence SNe have much larger weights than GRBs in the fitting.

\begin{table}
\centering
\caption{\small{The best-fitting parameters of $\Lambda$CDM and $\omega$CDM models with different GRB subsamples.}}\label{tab:cosmological parameters_G}
\arrayrulewidth=1.0pt
\renewcommand{\arraystretch}{1.3}
{\begin{tabular}{cccc} 
\hline\hline 
& \multicolumn{1}{c}{$\Lambda$CDM}  & \multicolumn{2}{c}{$\omega$CDM}\\
\cline{2-4}
GRBs & $\Omega_{\rm M}$  & $\Omega_{\rm M}$     & $\omega$   \\\hline
low-$z$     &0.306$^{+0.105}_{-0.133}$ &0.356$^{+0.132}_{-0.132}$  &$<-1.125$	\\
high-$z$     &0.195$^{+0.069}_{-0.097}$ &0.217$^{+0.103}_{-0.107}$  &$<-0.963$	\\
all-$z$ &0.280$^{+0.049}_{-0.057}$ &0.345$^{+0.059}_{-0.060}$  &$<-1.414$ \\
\hline
\end{tabular}}
\end{table}

\begin{table}
\centering
\caption{\small{The best-fitting parameters of $\Lambda$CDM and $\omega$CDM models with the combination of SNe Ia and different GRB subsamples.}} \label{tab:cosmological parameters_SG}
\arrayrulewidth=1.0pt
\renewcommand{\arraystretch}{1.3}
{\begin{tabular}{cccc} 
\hline\hline 
& \multicolumn{1}{c}{$\Lambda$CDM}  & \multicolumn{2}{c}{$\omega$CDM}\\
\cline{2-4}
SNe+GRBs & $\Omega_{\rm M}$  & $\Omega_{\rm M}$     & $\omega$   \\\hline
SNe+low-$z$     &0.279$^{+0.008}_{-0.008}$ &0.345$^{+0.058}_{-0.053}$  &$-1.166^{+0.173}_{-0.148}$	\\
SNe+high-$z$     &0.278$^{+0.008}_{-0.008}$ &0.314$^{+0.063}_{-0.061}$  &$-1.083^{+0.174}_{-0.137}$	\\
SNe+all-$z$    &0.278$^{+0.008}_{-0.009}$ &0.336$^{+0.055}_{-0.050}$  &$-1.141^{+0.156}_{-0.135}$ \\
\hline
\end{tabular}}
\end{table}

\section{Discussions and conclusions}\label{sec:conclusion}

In this paper, we have investigated the possible redshift dependence of Combo-relation using deep learning. We constructed a network which combines RNN and BNN, and trained the network using the Pantheon compilation of SNe Ia. The distance-redshift relation of SNe was reconstructed model-independently up to the highest redshift of GRBs. By dividing GRB sample into two subsamples with redshift boundary $z=2$, the Combo-relation was tested separately for low-$z$ and high-$z$ subsamples. The results show that the Combo-relation of low-$z$ GRBs is consistent with that of high-$z$ GRBs within $1\sigma$ confidence level. Thus the tightness and redshift-independence of Combo-relation is confirmed. We further calibrated the distance of the full GRB sample and the two subsamples using Combo-relation. The calibrating method is model-independent so that the calibrated GRBs can be used directly to constrain cosmological models. For $\Lambda$CDM model, the constraints with GRBs alone are consistent with the result of Pantheon sample within 1$\sigma$ uncertainty. For the $\omega$CDM model, however, the EoS of dark energy $\omega$ couldn't be tightly constrained by GRBs alone. The accuracy of constraint on $\omega$CDM model is improved when adding GRBs to SNe.

\citet{Muccino et al.:2021} have already used the same GRB sample to test the redshift dependence of Combo-relation. They divided GRBs into seven subsamples. In each subsample, GRBs are approximately at the same redshift, hence have a constant distance. Therefore, the left-hand-side term of equation (\ref{eq:Combo}), $\log\left[L_0/(\rm erg/s)\right]$, can be replaced by $\log\left[F_0/(\rm erg/cm^2/s)\right]$, and the constant term $\log\left[4\pi d_L^2/{\rm cm^2}\right]$ can be absorbed in to the $\log\left[A/(\rm erg/s)\right]$ term on the right-hand-side of equation (\ref{eq:Combo}). The slope parameters $\gamma$ obtained in each subsample is independent of cosmological model. \citet{Muccino et al.:2021} found that the values of $\gamma$ for seven subsamples are not redshift-evolving, with the mean value $\gamma=0.90\pm 0.13$, and an intrinsic scatter $\sigma_{\rm int}=0.28\pm 0.03$. These results are consistent with ours. As is mentioned in the introduction, at high redshift it is difficult to find sufficient GRBs approximately at the same redshift. The highest redshift subsample that \citet{Muccino et al.:2021} used to obtain $\gamma$ has the mean redshift $z=2.69$. There is a large redshift gap between $z=2.69$ and the highest redshift of GRB ($z=9.4$). In this redshift gap the redshift dependence of $\gamma$ has not been tested. By fixing $\gamma$ to the mean value, \citet{Muccino et al.:2021} calculated the intercept parameter $\log [A/{\rm (erg/s)}]$ by using SNe to anchor the distance of four nearest GRBs. They obtained the average value $\log [A/{\rm (erg/s)}]=49.54\pm 0.21$, which is also consistent with our result. Although this method can minimize the use of SNe, it strongly depends on the accuracy of a handful of SNe and GRBs that are used to anchor the distance. In addition, \citet{Muccino et al.:2021} assumed that the intercept parameter $\log [A/{\rm (erg/s)}]$ is a constant in the full redshift range, which couldn't be tested by the calibrating method itself. In comparison, we tested the redshift dependence of the slope and intercept parameters simultaneously using the full sample of SNe and GRBs. Our calculating method strongly depends on the validity of distance-redshift relation of SNe, but is less affected by the possible outliers of data points.

\section*{Acknowledgements}
This work has been supported by the National Natural Science Fund of China Grant Nos. 11603005, 11775038 and 11947406.

\section*{Data Availability}
The Pantheon dataset is available at \textcolor{blue}{https://archive.stsci.edu/prepds/ps1cosmo/index.html}. The GRB dataset is taken from \citet{Muccino et al.:2021}, which is available at \textcolor{blue}{https://iopscience.iop.org/article/10.3847/1538-4357/abd254}.

\bsp

\label{lastpage}

\end{document}